\documentstyle[11pt,newpasp,epsf,twoside]{article}
\def\ace{\alpha_{\rm CE}}
\def\Halpha{H$\alpha$}
\def\HST{{\it HST}}
\def\IUE{{\it IUE}}
\def\kms{{\>\rm km\>s^{-1}}}
\def\ltabout{\la}
\def\OIII{\hbox{O~$\scriptstyle\rm III$}}
\def\subsun{_{\odot}}

\begin{document}

\title{Binarity of Central Stars of Planetary Nebulae}
\author{Howard E. Bond}
\affil{Space Telescope Science Institute, 3700 San Martin Dr., Baltimore, MD 
21218 USA}

\begin{abstract}
I list the 16 planetary nebulae (PNe) known to contain close-binary nuclei,
and show that the nebulae generally have axisymmetric structures,
including elliptical, bipolar, or ring morphologies. The orbital periods range
from 2.7~hr to 16~days, and close binaries constitute $\approx\!10\%$ of all
central stars. Since the known binaries were found mainly from photometric 
variability, which depends on heating effects at very small stellar 
separations, radial-velocity surveys will be necessary to find the large
predicted population of binary nuclei with periods of about 10-100 days. Other
PN phenomena that may arise from binary-star interactions include jets and
point-symmetry, the periodically spaced arcs revealed by \HST\/ in the faint
halos around several PNe and proto-PNe, and the existence of PNe in globular
clusters. There is thus considerable circumstantial evidence that binary-star
processes play a major role in the formation and shaping of many or even most
PNe. 
\end{abstract}

\section{Introduction}

The subtitle of this paper could be ``Do close companions strongly influence,
or even {\it cause}, the ejection of most planetary nebulae?''  The answer to
this question may well be  ``yes''---I certainly think it is---but as we will
see the evidence is still incomplete. 

I want to acknowledge the contributions of my collaborators in various
portions of this work, including D. Alves, R. Ciardullo, M. Cohen, L. Fullton,
M.~Livio, K. Schaefer, M. Sipior, H. Van Winckel, and C.-Y. Zhang. 

\section{PNe with Known Close-Binary Nuclei} 

\looseness=-1
I begin by considering those planetary nebulae (PNe) where we {\it know\/} the
central star is a close binary, because it (a)~exhibits periodic photometric
variability (due to actual stellar eclipses, or to heating
effects on a main-sequence companion of the hot nucleus), (b)~is a
short-period spectroscopic binary, or (c)~has a composite spectrum along with
evidence of interaction between the two stars.  There are 12 objects currently
known in the first class, and one in the second (NGC~2436, first discovered as
a spectroscopic binary, which also shows episodic periodic light variations).
The third class contains three ``Abell~35-type'' planetary-nebula nuclei
(PNNi), in which a cool, rapidly rotating optical star has a hot companion
revealed in \IUE\/ UV spectra. All three A~35-type PNNi show low-amplitude
light variations with periods of a few days (e.g., 5.9~days for LoTr~5---Bond
\& Livio 1990; Strassmeier, Hubl, \& Rice 1997), but these arise from
starspots on the rotating cool stars, and the true orbital periods remain as
yet unknown.  The rapid rotation of the optical companions is believed to
result from accretion of angular momentum from the wind of the AGB progenitor
of the star that is now the hot PNN (e.g., Jeffries \& Stevens 1996; Gatti
et~al.\ 1997, 1998). All 16 objects are listed in Table~1.


\medskip

\def\p{\phantom{1}}
\def\z{\phantom{3}}

\centerline{\bf Table 1}
\centerline{\bf Planetary Nebulae with Close-Binary Nuclei}
\vskip-0.2in
\begin{center}
\tabcolsep=1.25em
\begin{tabular}{llcl}
\tableline
\noalign{\smallskip}
\hfil Planetary&\hfil Central& Period&\hfil Binary\\
\hfil Nebula&\hfil Star& (days)&\hfil Type\\
\noalign{\smallskip}
\tableline
\noalign{\smallskip}
Abell 41& MT Ser& 0.113&Reflection\\
DS 1& KV Vel& 0.357&Reflection\\  
Hf 2-2& (MACHO var.)& 0.399&Reflection\\
Abell 63& UU Sge& 0.465&Eclipsing\\
Abell 46& V477 Lyr& 0.472&Eclipsing\\
HFG 1& V664 Cas& 0.582&Reflection\\
K 1-2& VW Pyx& 0.676&Reflection\\
Abell 65& \hfil$\dots$& 1.00\z&Reflection\\
HaTr 4& \hfil$\dots$& 1.74\z&Reflection\\
(Tweedy 1)& BE UMa& 2.29\z&Eclipsing\\
SuWt 2& \hfil$\dots$& 2.45\z&Eclipsing\\
Sp 1& \hfil$\dots$& 2.91\z&Reflection\\
NGC 2346& V651 Mon& \llap{1}5.99\z&Spectroscopic\\
Abell 35& BD $-22\deg3467$& $\dots$&\IUE\ composite\\
LoTr 1& \hfil$\dots$& $\dots$&\IUE\ composite\\
LoTr 5& HD 112313& $\dots$&\IUE\ composite\\
\noalign{\smallskip}
\tableline
\end{tabular}
\end{center}

\medskip


Table~1 is similar to the one I presented at Asymmetrical Planetary Nebulae~I
(Bond 1995), with two main additions: the period of the eclipsing nucleus of
SuWt~2 is now established as 2.45~days, from my recent CCD photometry at Cerro
Tololo; and a new short-period reflection binary has been discovered in Hf~2-2
by the MACHO collaboration (Lutz et~al.\ 1998). 

As noted previously (Bond 1995 and references therein), the fraction of
detectable close binaries among randomly selected PNNi is of order 10--15\%.
(This statement is based on the fact that the 13 variables of Table~1 were
found in photometric surveys of somewhat more than 100 PNNi). In a recent
analysis of the MACHO Galactic bulge database, Lutz et~al.\ (1998) found one
close binary (Hf~2-2, mentioned above) out of 22 random PNNi that fell within
the MACHO survey area, giving a close-binary fraction of $4.5\pm4.5$\%. This
is in tolerable agreement with the figure quoted above, given the small-number
statistics. 

The fraction of known very close binaries among PNNi is remarkably high.
Moreover, the photometric search technique\footnote{Large-scale synoptic
photometric surveys are, unfortunately, becoming almost impossible at the U.S.
national observatories due to closures of small telescopes} reveals only
binaries with short periods ($\la$ a~few days), since it relies on eclipses
and heating effects.  It is thus quite possible that the short-period PNNi are
only the tip of an iceberg of a substantial population of longer-period
binaries. 

\looseness=-1
But do we expect to find PNNi with longer binary periods? The binaries in
Table~1, with periods of 2.7~hr to 16~days, are so close that the nebulae {\it
must\/} have been ejected through common-envelope (CE) interactions (see Bond
\& Livio 1990; Iben \& Livio 1993; Sandquist et~al.\ 1998; and references
therein). Several groups (e.g., de~Kool 1990; Yungelson, Tutukov, \& Livio
1993; Han, Podsiadlowski, \& Eggleton 1995) have performed simulations in
which they evolve a population of primordial binaries through the stage of
formation of a red giant and, if the system is close enough, the subsequent
ejection of a CE accompanied by a spiral-down of the orbit. The outcome of
these syntheses depends on the efficiency parameter $\ace$, which is the
fraction of the orbital gravitational energy that goes into ejecting material
from the envelope into space. For $\ace\approx1$ a substantial fraction of the
resulting binaries will have periods of 10 to several hundred days.  In that
case, since the fraction of {\it short-period\/} binaries (with periods of
$\approx\!0.1$ to 10~days) is already $\approx\!10\%$, the {\it total\/}
fraction of binaries could be quite large.  On the other hand, if the crucial
$\ace$ parameter is small, $\la0.1$, then the known short-period fraction
could represent substantially all of the binaries that have emerged from CE
interactions (other than those binaries that have actually coalesced). 

\looseness=-1
Thus the most crucial next observational step in this subject would be a
radial-velocity survey of PNNi for binaries with $10 \,{\rm days} \ltabout P
\ltabout 100\,\rm days$.  The binary fraction in this range would provide a
crucial constraint on the value of $\ace$, and would tell us whether CE 
ejections are the primary way of making PNe.

\section{Morphologies of PNe Ejected from Common Envelopes} 

I now consider the morphologies of PNe which are {\it known\/} to have
close-binary nuclei. Previously published papers have addressed this subject
in detail (Bond \& Livio 1990; Walton, Walsh, \& Pottasch 1993; Pollacco \&
Bell 1997), and generally it is found that such PNe are axisymmetric, showing
either elliptical or more pronounced butterfly or bipolar shapes.  I showed
a number of new ground- or space-based images at the workshop, but this
printed version allows space for only a few examples of the most interesting
classes of phenomena. 

{\bf Non-spherical morphologies}. These are ubiquitous among PNe known to have
close-binary nuclei. Fig.~1, for example, shows ground-based images of the
newly discovered (Liebert et~al.\ 1995) PN around the eclipsing binary BE~UMa,
and of the PN A~41 whose central star is a binary with the extremely short
period of 2.7~hr (Grauer \& Bond 1983). BE~UMa, where we know we view the
binary orbit edge-on, shows a rectangular structure (probably an edge-on
cylinder), and an apparent wind-blown bubble along an axis perpendicular to
the presumed orbital plane on the upper left side.  A~41 likewise shows bright
structures on either side of the nucleus, again indicating a cylindrical
morphology, and again with incipient wind-blown bubbles along the axis.  The
bipolar structure is even more pronounced in recent {\it Hubble Space
Telescope\/} (\HST\/) images of NGC~2346 (not shown), which has a classical
butterfly morphology. 

\begin{figure}[hbt]

\hbox to \textwidth{%
 \null\hfil
 \hbox{\epsfysize=2.2in \epsffile{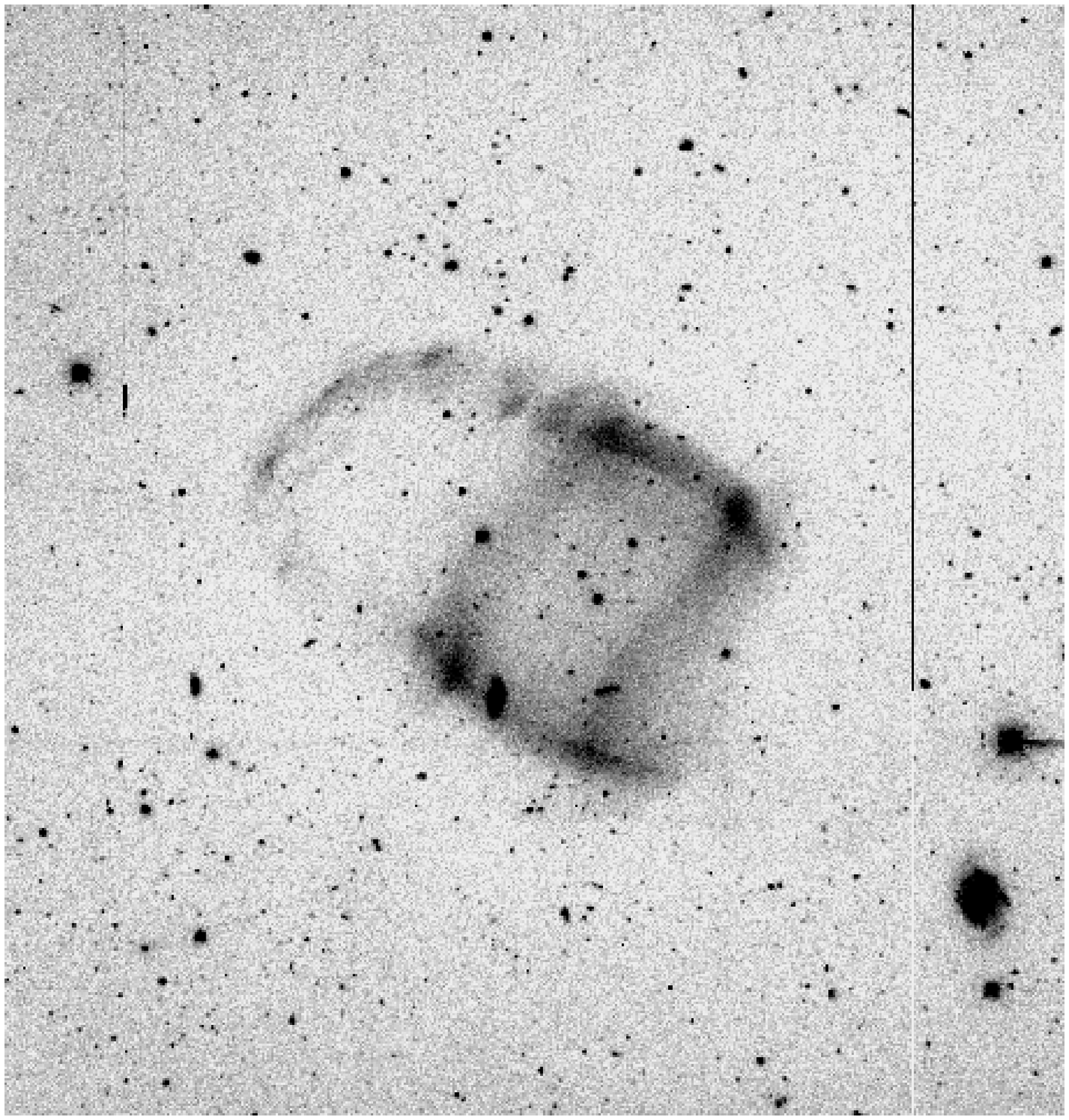}}
 \hfil
 \hbox{\epsfysize=2.2in \epsffile{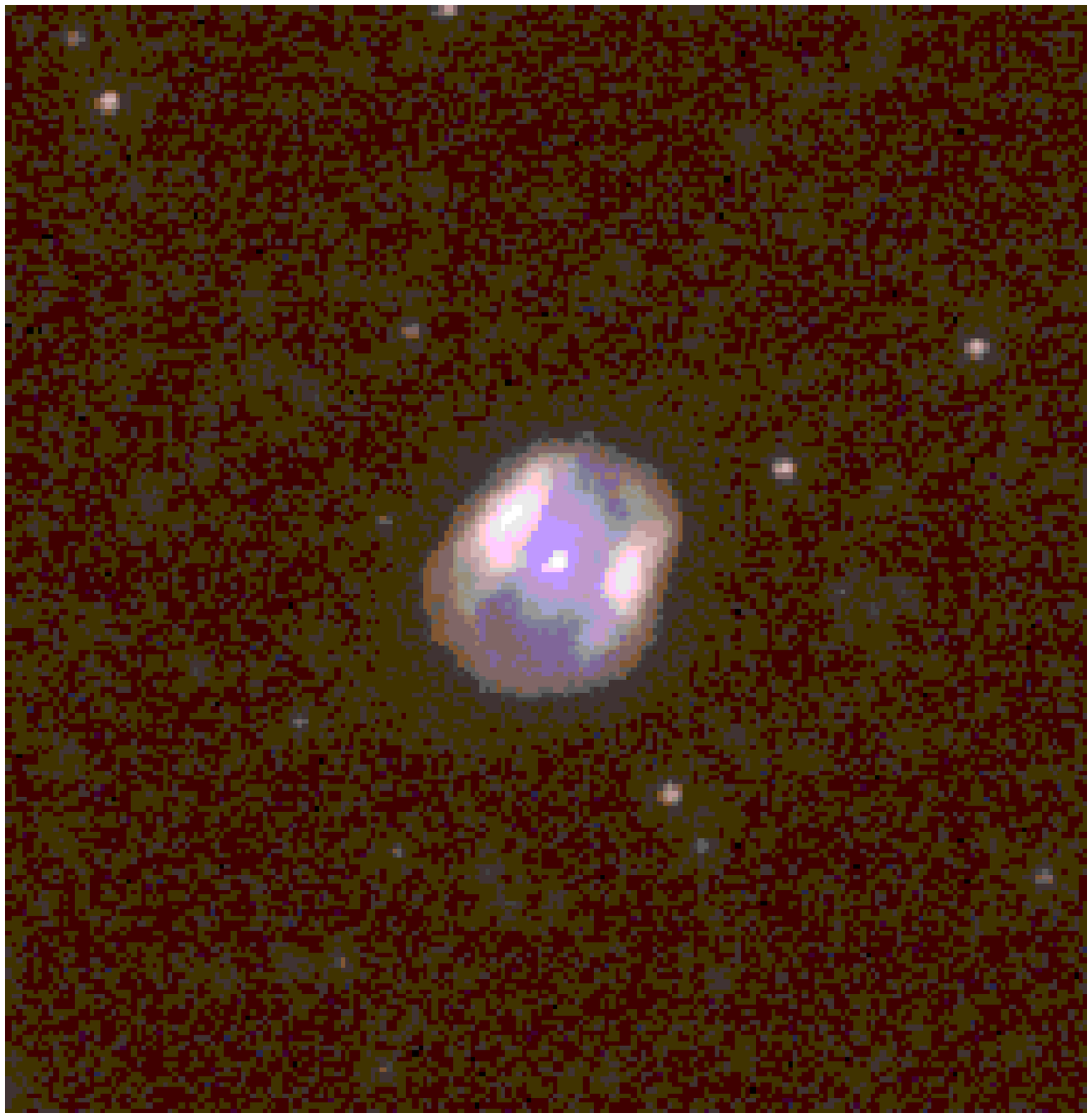}}
 \hfil
}
\caption[]{Ground-based images of the BE~UMa PN ({\bf left}) and A~41 ({\bf
right}).  The BE~UMa image was taken in [\OIII] $\lambda$5007 with the Kitt
Peak 4-m, and the A~41 image is the sum of [\OIII] and \Halpha, obtained with
the Cerro Tololo 0.9-m. Note the axisymmetric structure in both images, with
wind-blown balloons at the poles.} 
\end{figure}

{\bf Ring nebulae}. By contrast, Sp~1 has a nearly circular morphology, as
shown in the left-hand image of Fig.~2. We argued (Bond \& Livio 1990) that
Sp~1 is in fact a ring or cylinder (rather than a spherical structure), seen
almost pole-on. This was in accord with the low photometric amplitude of the
reflection effect in the central binary, which would indeed be small in a
binary viewed close to pole-on. If so, we would expect to see other PNe around
binary nuclei viewed at less extreme angles as ellipses rather than circles.
Gratifyingly, SuWt~2 (Fig.~2, right-hand side) does show an elliptical shape,
and the central star is an {\it eclipsing\/} binary.  These ``wedding-ring''
PNe have extremely high ``density contrasts'' between their equatorial and
polar regions. 

\begin{figure}[hbt]

\hbox to \textwidth{%
 \null\hfil
 \hbox{\epsfysize=2.55in \epsffile{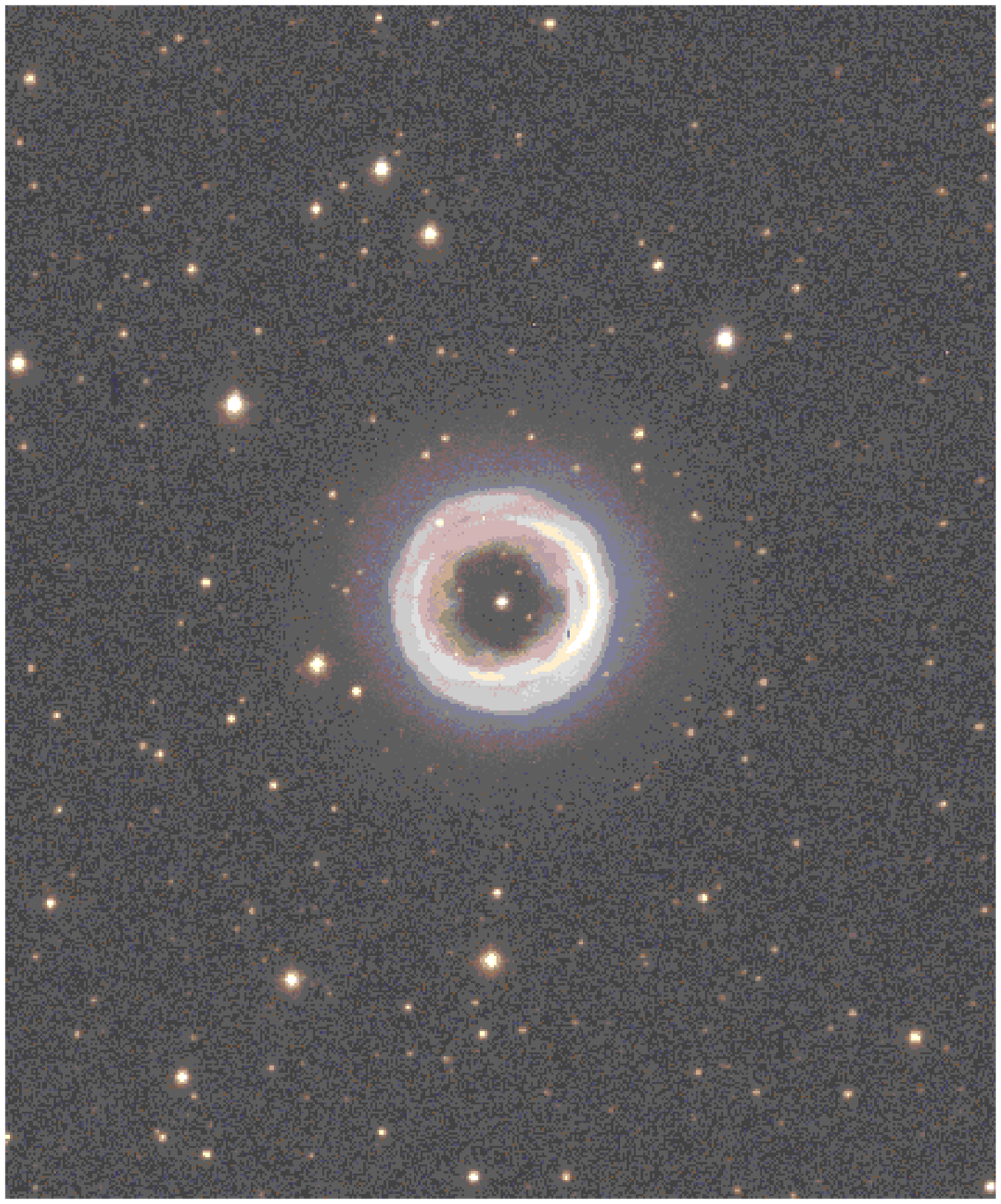}}
 \hfil
 \hbox{\epsfysize=2.55in \epsffile{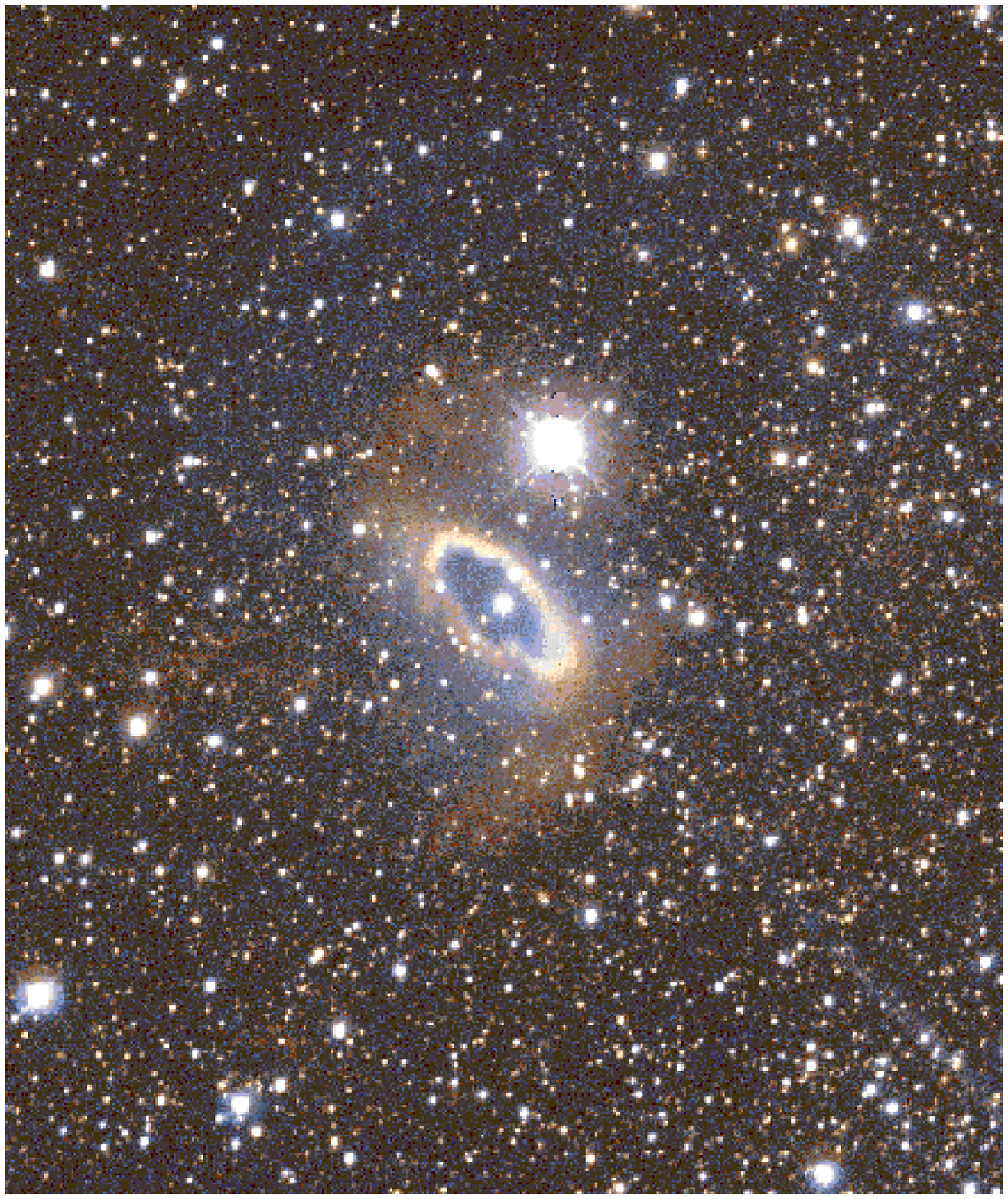}}
 \hfil
}
\caption[]{Ground-based [\OIII]+\Halpha\ images of Sp~1 ({\bf left}) and
SuWt~2 ({\bf right}), obtained with the Cerro Tololo 0.9-m. Both PNe appear to
be rings with very low ratios of height to radius. The ring in SuWt~2 is
viewed close to edge-on, and the central star is an eclipsing binary.} 
\end{figure}

\looseness=-1
In summary, PNe known to contain close-binary nuclei are virtually all
axisymmetric, with varying degrees of density contrasts. An unsolved question,
of course, is whether the opposite is true, i.e., are {\it all\/}
non-spherical PNe ejected from binaries? This hypothesis was debated at length
at this workshop, but although it remains (in my opinion) highly plausible, it
is still observationally unproven. 

\section{PNe with Resolved Visual-Binary Nuclei} 

Wide binaries with initial periods less than $\approx\!10^3$~days, but still 
wide enough for the primary to evolve to red-giant dimensions, will undergo
a CE interaction and spiral down to much shorter periods, or coalesce (e.g.,
Yungelson et~al.\ 1993). Those with larger initial separations will not enter
into a CE, and in fact their separations will increase somewhat when the
primary star ejects a PN\null. The widest of these latter
binaries may be seen as
resolved visual binaries. 

We (Ciardullo et~al.\ 1999) carried out a \HST\/ snapshot survey of PNNi with
the Wide Field Camera~2, in order to search for the predicted population of
resolved visual binaries in PNe. Out of 113 PNNi, a dozen probable or possible
resolved nuclei were found. These binaries are extremely useful for finding
accurate ($\pm20\%$) distances to PNe, through main-sequence fitting of the
companions. The results, as shown in Ciardullo et~al., reveal that the
previous statistical distances for PNe have been systematically overestimated,
by up to a factor of $\sim\!2$. However, in the context of this presentation,
these very wide companions probably do not have major effects on the PN
morphology.  See, however, Soker (1999), who argues that the PN ejection
process may be sufficiently leisurely for the orbital acceleration of the PNN
during the ejection to lead to somewhat non-spherical morphologies. 
                  
\section{Other PN Phenomena that May Require Binaries}

There are several other phenomena seen in PNe that may arise from binary-star 
interactions, although in general this hypothesis remains as yet unproven.

\subsection{Jets and Point-Symmetry}

\looseness=-1
Jets, in all other astrophysical settings, are believed to result from 
outflows that are collimated by accretion
disks. Thus it is a surprise to find jets in PNe, where one usually thinks of 
mass loss rather than accretion.  Fig.~3 (left) shows diametrically 
opposed jets on either side of the nucleus of K~1-2.
In a PN nucleus, formation of a disk probably
requires mass transfer between binary companions, and thus K~1-2, where we 
know the PNN is a close binary, is instructive. A related phenomenon is
point-symmetry, which is seen in a number of PNe; a spectacular example is 
shown in the \HST\/ image of NGC~5307, shown in Fig.~3 (right), in which 
nearly every blob and spiral-like feature has a counterpart on the opposite 
side of the PNN\null.
Point-symmetric structure could arise from an episodic jet from a
precessing disk (e.g., Livio \& Pringle 1997), but it is not yet known whether 
the nuclei of NGC~5307 or similar PNe are binaries.

\begin{figure}[hbt]

\hbox to \textwidth{%
 \null\hfil
 \hbox{\epsfysize=2.4in \epsffile{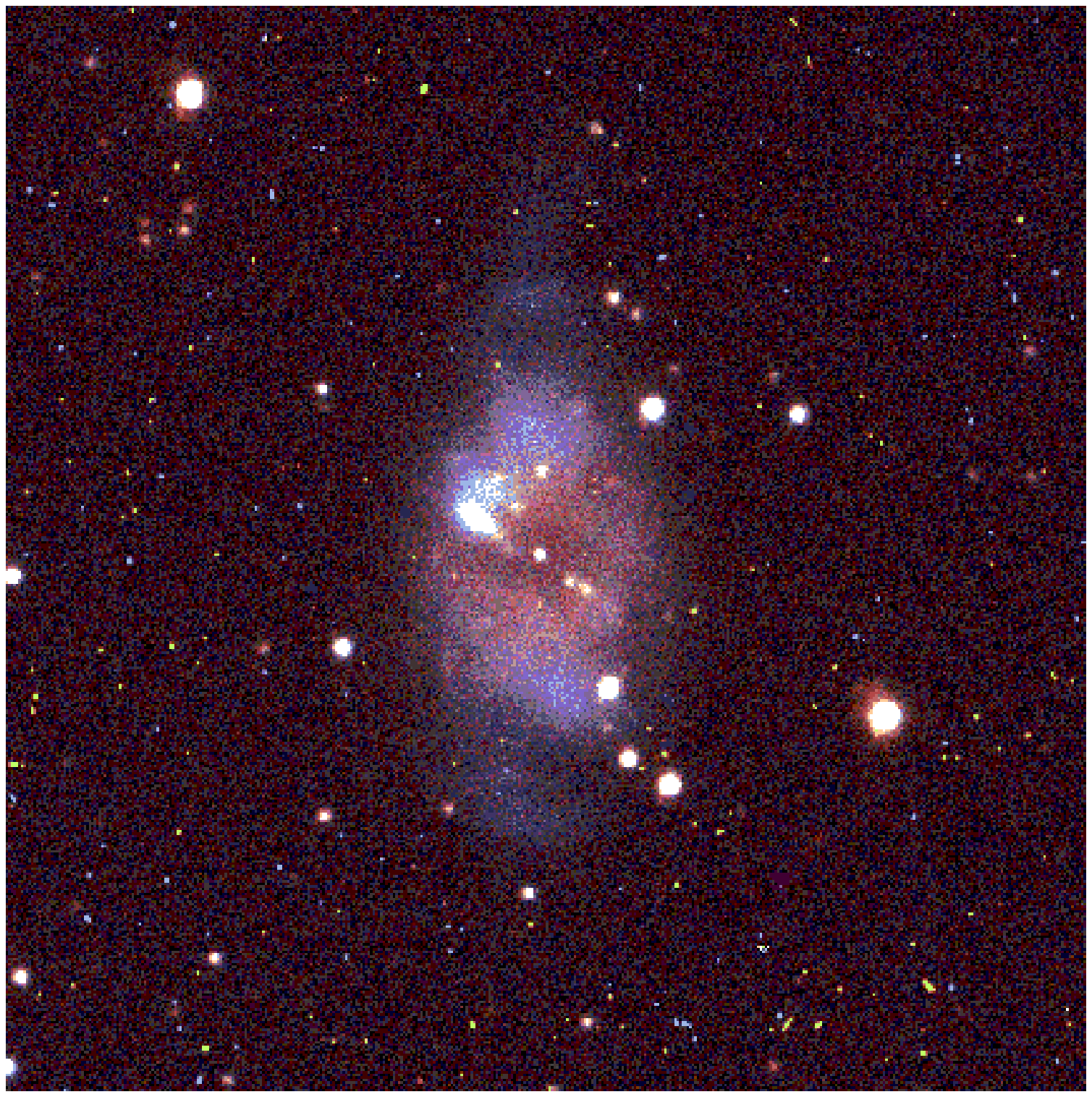}}
 \hfil
 \hbox{\epsfysize=2.4in \epsffile{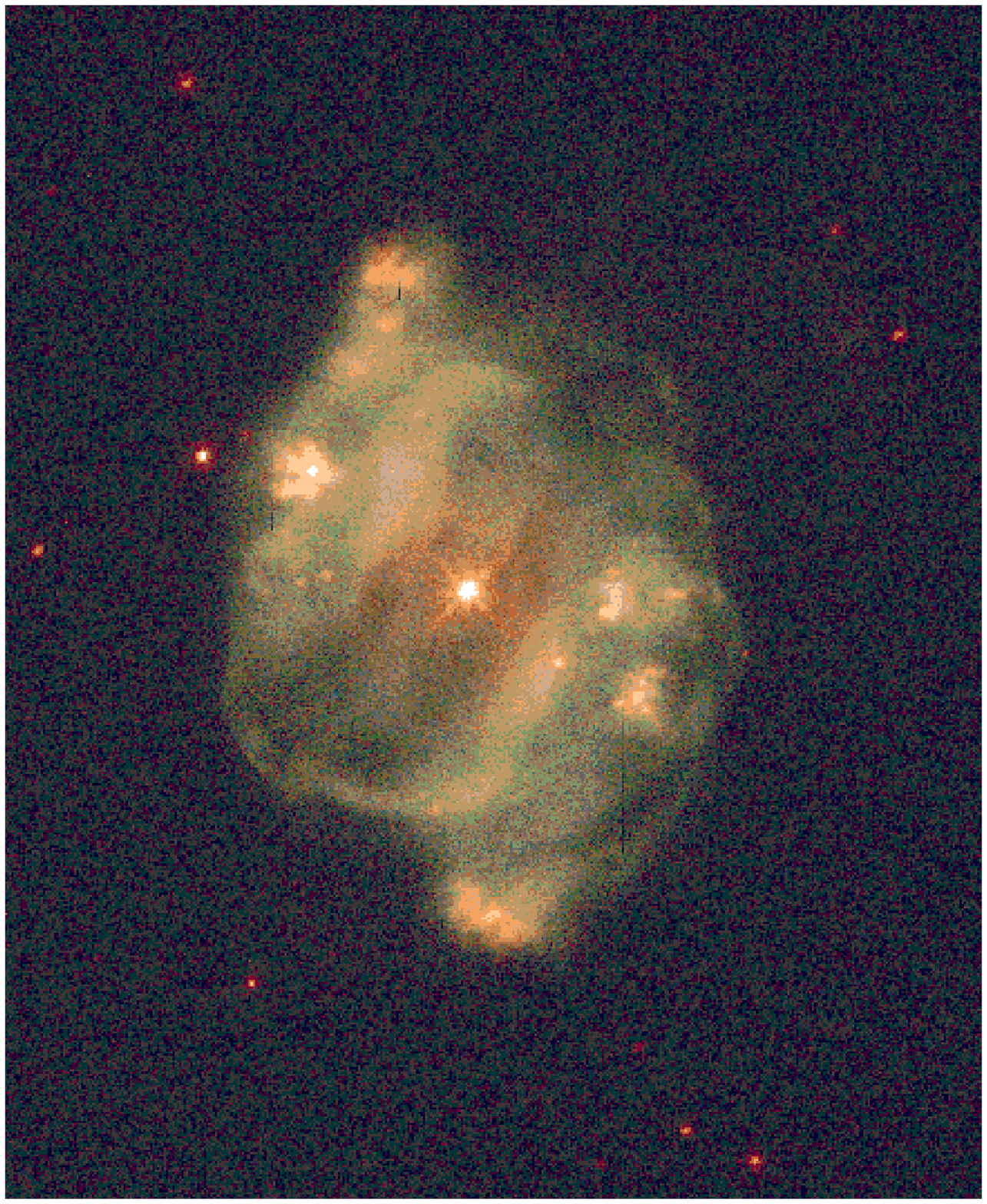}}
 \hfil
}
\caption[]{({\bf Left}) Ground-based [\OIII]+\Halpha\ image of K~1-2,
obtained with the Cerro Tololo 0.9-m. Note jets on either side of the 
close-binary nucleus. ({\bf Right}) Broad-band ($V$+$I$) \HST\/ WFPC2 image of 
the point-symmetric PN NGC~5307, from the snapshot survey of Bond \& 
Ciardullo.}
\end{figure}

\subsection{Concentric Periodic Rings}

\HST\/ imaging 
has revealed a remarkable new PN phenomenon: quasi-periodic concentric
arcs in the halos of several bipolar PNe and proto-PNe. Examples include the
PNe NGC~7027 (Fig.~4, left) and NGC~6543 (Fig.~4, right), and the proto-PNe
CRL~2688 (the ``Egg'' Nebula; Sahai et~al.\ 1998) and IRAS
17150$-$3224 (Kwok, Su, \& Hrivnak 1998). In Fig.~4, I have processed both
\HST\/ images through unsharp masking to enhance the visibility of the faint 
rings around the bright inner nebulae. More recent \HST\/ images reveal
quasi-periodic {\it linear\/} 
features in the ``Red Rectangle'' (HD~44179; Cohen
et~al.\ 1999). 

\begin{figure}[hbt]

\hbox to \textwidth{%
 \null\hfil
 \hbox{\epsfysize=2.4in \epsffile{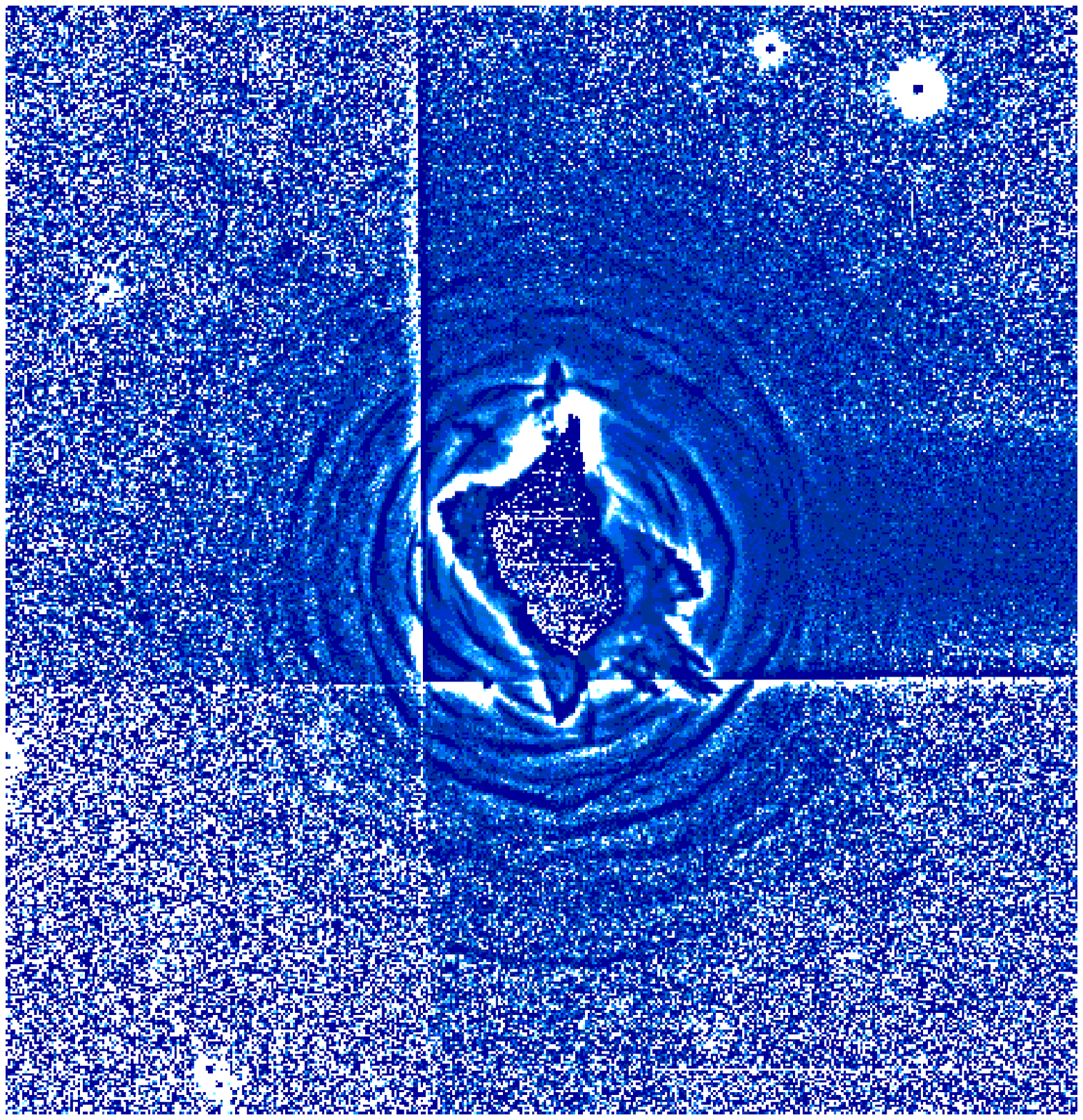}}
 \hfil
 \hbox{\epsfysize=2.4in \epsffile{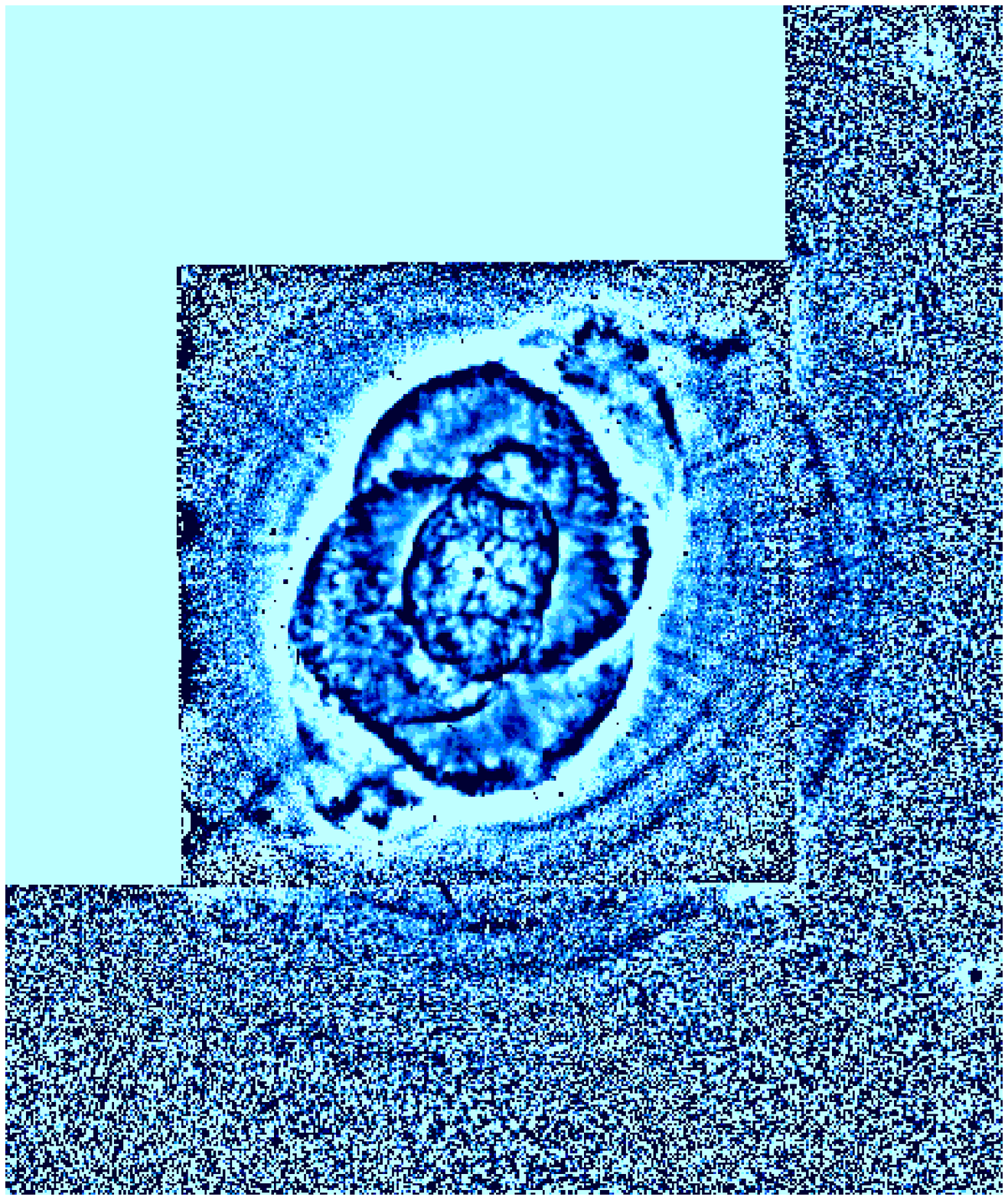}}
 \hfil
}
\caption[]{\HST\/ \Halpha\ images of NGC~7027 ({\bf Left}) and NGC~6543
({\bf Right}).  Both images have been processed with unsharp masking, and 
reveal the faint, nearly circular and nearly regularly spaced system of arcs 
surrounding these nebulae.}
\end{figure}

In NGC~7027, the characteristic spacing of the arcs is $\Delta r\simeq3''$.
If these features are due to modulated mass loss, the time interval between
ejections is

$$\Delta t \simeq 575 \, {\rm yr}
\left({\Delta r\vphantom{d}}\over{3''\vphantom{p}}\right)
\left({v_{\rm exp}\vphantom{d}}\over{20 \kms\vphantom{p}}\right)^{-1}
\left(d\over{800 \,\rm pc}\right).$$

The rings suggest that (a)~the early outflow from the AGB progenitor was
spherical, and later became aspherical when the densest portion of the PN was
ejected, and (b)~the early, spherical outflow was {\it episodic}. 

\looseness=-1
The putative timescale of $\sim\!600$~yr is too short to be due to thermal
pulses in the AGB progenitor, and too long to be due to envelope pulsations.
Other suggested origins of the periodicity include (a)~a long-term amplitude
modulation of envelope pulsations (Icke, Frank, \& Heske 1992; Sahai et al.\
1998); (b)~the lifetime of individual giant convective cells (Schwarzschild
1975); (c)~an instability in the radiation-driven gas/dust outflow (Morris
1992; Deguchi 1997) producing density waves in the dust; or (d)~the influence
of a binary companion.  The last option is of course the one of interest in
the present context. Harpaz, Rappaport, \& Soker (1997) proposed that the
rings could arise from the influence of a companion (even of planetary mass)
in an eccentric orbit, but the number of PNe showing the phenomenon is now
large enough that it seems unlikely that so many of the nuclei would have
eccentric companions in 600-yr orbits. However, Mastrodemos \& Morris (1999)
have argued more recently that the rings could originate from spiral shocks
produced by binary companions in circular orbits, and the predicted statistics
for the frequency of occurrence of such companions (e.g., Yungelson et~al.\ 
1993) seem in rough accord with the statistics for the periodic rings. 

\subsection{PNe in Globular Clusters}

\looseness=-1
Here I ask the question ``why are there PNe in globular clusters?''\ because I
will show that there should not be any. Bond \& Fullton (1999) have searched
for post-AGB stars in the halo of M31, and show from the star counts that the
post-AGB transition time through types F and~A is $\approx\!25,000$~yr for
these remnants of low-mass ($\sim\!0.8M\subsun$) stars. Since the timescale for
dissipation of a PN is thus shorter than the post-AGB evolutionary timescale
for halo stars, {\it any PN should be gone by the time the central star
becomes hot enough to ionize it}, and this argument should apply in globular
clusters (GCs)
as well as in galactic halos. However, Jacoby \& Fullton (see Jacoby
et~al.\ 1997), in a complete survey of Milky Way GCs,
found two new PNe, bringing the total number known in GCs to four. 

One way to resolve the puzzle is to suppose that the PNe that are seen in GCs
are descended from binary stars, in which the mass of one component was raised
either by mass transfer, or by stellar coalescence. In this case, the remnant
would have a higher mass than remnants from single stars, and would evolve
much more rapidly across the HR diagram, thus allowing it to ionize the PN
while there is still time. 

In order to test this hypothesis, we (Alves, Bond, \& Zurek 1999) used \HST's 
WFPC2
to monitor the central star of the PN K~648 in the GC M15. In addition to the
post-AGB lifetime argument summarized above, the known luminosity of the K~648
PNN places it on a post-AGB track of $0.58M\subsun$ (Bianchi et~al.\ 1995),
considerably more massive than normal white dwarfs in GCs (Richer et~al.\
1997), again suggesting that the star has experienced a mass augmentation at
some point in its past. Unfortunately, however, the star did not display any
variability in our \HST\/ photometry, periodic or otherwise, on timescales of
45~min to about 10~days. But this does not rule out the hypothesis of
binarity, since the PNN could still be a binary that is viewed pole-on, a
binary with an orbital separation too large for significant heating effects,
or a coalesced binary. 

\section{Conclusions}

I would like to conclude by asking the reader to consider two syllogisms:

\vskip -0.2 in

\def\medstrut{\hbox{\vrule height 0.75em depth 0.38em width 0pt}}
\def\smallstrut{\hbox{\vrule height 0.3em depth 0em width 0pt}}
\def\tablestrut{\hbox{\vrule height 1.0em depth 0.3em width 0pt}}
\def\ul#1{$\underline{\hbox{\vphantom{y}#1}}$} 

$$\vbox{\offinterlineskip
\halign{\medstrut
  #\hfil\tabskip=2em plus 1fil minus 1.25em&  
  #\hfil\tabskip=0pt\cr                     
\hfil \bf\ul{Planetary Nebulae} & \hfil \bf\ul{Ducks} \cr
\noalign{\smallstrut}
1. Binaries eject axisymmetrical PNe. & 1. Ducks quack.      \cr
2. Most PNe are axisymmetrical.             & 2. I quack.              \cr
3. Therefore most PNe are ejected     & 3. Therefore I am a duck. \cr
\phantom{3. }from binaries. &                                      \cr
}}$$

\vskip -0.1 in

The syllogism on the left about PNe seems plausible, especially in the light
of the circumstantial evidence I have presented in this paper. However, we
should beware of the logical flaw in this syllogism, whose lack of
rigor\footnote{See, however, {\tt http://us.imdb.com/Title?0091225}} is
exposed in the syllogism on the right by changing the argument to one about
ducks instead of PNe. 

As often happens, the poet (Robert Frost) has the last word: 

\smallskip

{\sl\obeylines

\centerline{We dance 'round in a ring and suppose,}
\centerline{But the Secret sits in the middle and knows.}
\par}

\smallskip

\acknowledgments
Support from NASA grant NAG~5-6821 and
STScI grants GO-6119, GO-6751, and GO-7297 is gratefully acknowledged.
Portions of the work reported were based on observations with the NASA/ESA 
{\it Hubble Space
Telescope}, obtained at the Space Telescope Science Institute, which is
operated by AURA, under NASA contract NAS~5-26555.

\end{document}